\documentclass[final, 5p, times, twocolumn, sort&compress]{elsarticle}
\usepackage{amsmath}
\usepackage{amssymb}
\usepackage{braket}
\usepackage{color}

\newcommand{\vb}[1]{\mathbf{#1}}            
\newcommand{\vh}[1]{\hat{\mathbf{#1}}}      
\newcommand{\cg}[6]{ 
        \Braket{#1 #2 \, #3 #4 | #5 #6}
}

\newcommand{\redmel}[3]{ 
        \Braket{#1 \| #2 \| #3}
}

\newcommand{\wtj}[6]{ 
        \begin{pmatrix}   % Wigner 3j symbol
                #1 & #2 & #3 \\
                #4 & #5 & #6 
        \end{pmatrix}
}

\newcommand{\wnj}[9]{ 
        \begin{Bmatrix}   % Wigner 9j symbol
                #1 & #2 & #3 \\
                #4 & #5 & #6 \\
                #7 & #8 & #9
        \end{Bmatrix}
}

\journal{Physics Letters B}

\begin{document}

\begin{frontmatter}

\title{Constraining the axial-vector X17 interpretation with ${}^{12}$C data}

\author[1]{Cornelis J.G. Mommers\corref{cor1}}
\ead{cmommers@uni-mainz.de}
\cortext[cor1]{Corresponding author.}
\author[1]{Marc Vanderhaeghen}
\affiliation[1]{
    organization={Institut für Kernphysik and PRISMA${}^+$ Cluster of Excellence, \\ Johannes Gutenberg-Universität}, 
    city={Mainz},
    postcode={D-55099},
    country={Germany}
}

\begin{abstract}
Recent findings of an unexpected, narrow resonance in the $e^+e^-$ decay spectra of excited states of $^8$Be, $^4$He and $^{12}$C by the ATOMKI collaboration have received considerable experimental and theoretical attention, whereby a new, 17-MeV vector-like or axial-vector-like boson termed X17 was conjectured as an explanation of the anomaly. Further analysis of all existing constraints disfavors a vector X17 scenario. 
For a similar analysis of the axial-vector scenario, a calculation of the reduced matrix element of a spin-dipole operator between the excited nuclear state ${}^{12}$C(17.23) and the carbon ground state is required. 
In the present work, we compute the aforementioned reduced matrix element under the assumption that the state ${}^{12}$C(17.23) is well represented by the $2s_{1/2}1p^{-1}_{3/2}$ particle-hole shell-model excitation of the ground state, as supported by experimental data. 
Within such a framework, our results indicate that, like the vector scenario, the axial-vector interpretation of X17 shows strong tensions with the other existing constraints on the nucleon coupling of a conjectured X17.  
\end{abstract}

\end{frontmatter}

\section{Introduction}
In a series of experiments the ATOMKI collaboration reported an anomalous, narrow resonance in the $e^+e^-$ decay spectra of excited states of $^8$Be, $^4$He and $^{12}$C \cite{Krasznahorkay:2015iga, Krasznahorkay:2015ijz, Krasznahorkay:2021joi, Krasznahorkay:2022pxs, Krasznahorkay:2023sax}. The observations have garnered significant interest and prompted multiple ongoing verification experiments, such as CCPAC \cite{Azuelos:2022nbu} and 
the PADME experiment \cite{Darme:2022zfw}. The first experiment to conclude, conducted at the VNU University of Science, recently reported results that were consistent with the original findings from the ATOMKI experiments \cite{Anh:2024req}.    

The ATOMKI collaboration conjectures that the anomalous signal does not originate from Standard-Model effects, but from a new, light vector or axial-vector boson with a mass of 17 MeV, referred to as X17 \cite{Alves:2023ree}. The experimental data are, in the case of ${}^8$Be and ${}^{12}$C, presented as ratios of partial widths of the excited nuclear state decaying to X17 relative to its photon decay, or, 
in the case of ${}^4$He, relative to its $e^+e^-$ E0 decay. The corresponding decay amplitudes can be related to several particle-physics models of X17 \cite{Feng:2016jff, Kozaczuk:2016nma} by assuming that the coupling of X17 to the proton and neutron are given in terms of the up and down quark couplings, $g_p = 2 g_u + g_d$ and $g_n = g_u + 2g_d$. The result is that $\Gamma_X / \Gamma_\gamma \propto |g_p \pm g_n|^2$, with a proportionality constant containing reduced matrix elements of nuclear multipole operators. Which combination is probed depends on the considered nuclear decay. The left-hand side of this equation is constrained by the ATOMKI experiments, which can then be translated into constraints on the right-hand side, $|g_p \pm g_n|$. 

The comprehensive analyses of Refs.~\cite{Barducci:2022lqd, Hostert:2023tkg} indicate that the vector interpretation of X17 is disfavored due to inconsistencies between different observations and preexisting bounds on the effective nucleon couplings. The same cannot yet be said about the axial-vector scenario due to two reasons. Firstly, for the $1^+ (18.15) \to 0^+(\text{g.s.})$ transition in ${}^8$Be, where the X17 signal was first observed, the leading partial wave of the vector decay is a $p$-wave, whereas the leading contribution for an axial-vector decay proceeds through an $s$-wave. Due to the extra two powers of momentum suppression in the vector versus axial-vector decay, the couplings of an axial-vector X17 are approximately two orders of magnitude smaller than the corresponding vector X17, which makes it easier to evade existing constraints in the axial-vector case.  Secondly, in the vector scenario the reduced matrix elements of the nuclear transition operators cancel in $\Gamma_X/\Gamma_\gamma$; they do not cancel in the axial vector scenario and must be explicitly computed. The resulting calculation has only been performed for the $^8$Be \cite{Kozaczuk:2016nma} and $^4$He \cite{Barducci:2022lqd} decays, but, so far, not for the $^{12}$C decay. Present results for the axial-vector X17 scenario appear to be consistent with each other, and with the preexisting bounds on the axial vector, up to uncertainties. However, without the ${}^{12}$C matrix elements, a definitive conclusion about the consistency of the axial-vector scenario cannot yet be drawn.

Explicitly, the relevant carbon decay is the E1 isovector decay,
\begin{equation*}
    {}^{12}\text{C}(17.23; 1^+, 1) \to {}^{12}\text{C}(\text{g.s.}; 0^+, 0) + \gamma/X17,
\end{equation*}
where the bracketed values indicate the energy of the excited state above the ground state (in MeV), spin-parity $J^P$, and isospin $T$ \cite{Kelley:2017qgh}. The decay to a photon is mediated by a dipole operator, and the decay to an axial-vector X17 is mediated by a spin-dipole operator, both defined further on. The main difficulty in determining the reduced nuclear matrix elements of these operators comes from the characterization of the excited carbon state.

The state ${}^{12}\text{C}(17.23)$ is broad and positioned at the onset of a dominant giant dipole resonance \cite{ALLAS1964122, Yamaguchi:1971ua}. 
Prior research, initially by Vinh-Mau and Brown, among others \cite{VINHMAU196289, GOSWAMI1963294, GILLET1964321}, and later by Lewis, Walecka and Donnelly \cite{PhysRevC.1.833, PhysRev.133.B849}, all consistently indicate that $^{12}\text{C}(17.23)$ is qualitatively well represented by a particle-hole shell-model state (1p1h), with only minor 2p2h contributions \cite{Grecksch1981}. The dominant configuration is $2s_{1/2} 1p_{3/2}^{-1}$ with negligible admixtures of other particle-hole configurations. If we approximate the $^{12}\text{C}(17.23)$ state as being entirely described by the $2s_{1/2} 1p_{3/2}^{-1}$ particle-hole excitation of the ground state, then the computation of the reduced matrix elements of the (spin-) dipole operator becomes tractable. The resulting calculation is presented in this work.

The outline of this Letter is as follows. In Sec.~\ref{sec:ph} we calculate the required reduced matrix elements of the dipole and spin-dipole operators mediating the $^{12}\text{C}(17.23) \to {}^{12}\text{C}(\text{g.s.}) + \gamma/X17$ decays, under the assumption that the excited carbon state is entirely described by the $2s_{1/2} 1p_{3/2}^{-1}$ particle-hole excitation of the ground state.
In Sec.~\ref{sec:disc} we compare our results to the existing limits constraining the nucleon couplings of X17 in case of the axial-vector scenario, and discuss the effectiveness of the 1p1h approximation. In Sec.~\ref{sec:outlook} we summarize our results and present an outlook.   

\section{$^{12}\mathrm{C}(17.23)$ particle-hole state, decay rates and transition operators}\label{sec:ph}

The spin-isopin-averaged isovector E1 decay rate via emission of an outgoing real photon or axial-vector X17 can be expressed in terms of the reduced matrix elements of the transverse electric or magnetic multipole operator as, 
\begin{align}
    &\Gamma(J_i T_i \to J_f T_f + \text{X17}/\gamma; \text{E1}) = \frac{2 |\vb{k}|}{\left( 2J_i + 1 \right)(2T_i + 1)} \nonumber \\
    &\times \sum_{M_{T_i}, M_{T_f}} \left\lvert  \wtj{T_f}{1}{T_i}{-M_{T_f}}{0}{M_{T_i}} \redmel{J_f T_f}{\vh{T}_1 \left(\left\lvert\vb{k} \right\rvert \right) \hat{T}_3}{J_i T_i} \right\rvert^2, \label{eq:decrate}
\end{align}
where $\vb{k}$ is the momentum of the outgoing X17 or photon with corresponding energies $E_X = \sqrt{|\vb{k}_X|^2 + m_X^2}$ and $E_\gamma = |\vb{k}_\gamma|$, and where $J_i$ and $J_f$ are the angular momenta of the initial and final nuclear states, respectively. The isospin of the initial (final) nuclear states are denoted by $T_i$ ($T_f$), with isospin projections $M_{T_i}$ ($M_{T_f}$) respectively. Unless stated otherwise all quantities in this section are given in the rest frame of the decaying nucleus. For an axial-vector X17 the one-body nuclear operator is given by an M1 operator $\vh{T}_1 = \vh{T}^\text{mag}_1$ and for the photon decay by an E1 operator $\vh{T}_1 = \vh{T}^\text{el}_1$, following Refs.~\cite{Barducci:2022lqd, Walecka:1995mi}). In both cases the isospin operator $\hat{T}_3$ selects the isovector decay channel. The reduced matrix element follows from the Wigner-Eckart theorem \cite{Suhonen:2007vjh},
\begin{align}
    &\Braket{\alpha_f; J_f M_f | T_{LM} | \alpha_i; J_i M_i } \nonumber \\
    &\quad = (-1)^{J_f - M_f} \wtj{J_f}{L}{J_i}{-M_f}{M}{M_i}
    \redmel{\alpha_f; J_f}{\vb{T}_L}{\alpha_i; J_i}.
\end{align}
Here, $T_{LM}$ is the $M$th component of the rank-$L$ spherical tensor $\vb{T}_L$, and $\alpha_i$ and $\alpha_f$ denote any other quantum numbers labeling the initial and final states. Throughout this Letter we follow the Condon-Shortley phase convention.

We parametrize the interaction of an axial-vector X17 with the nucleons by the effective Lagrangian, 
\begin{equation}
    \mathcal{L}_X = \sum_{N = n,p} g^A_N J^\mu_X X_\mu, \quad J^\mu_X = \bar{N} \gamma^\mu \gamma_5 N,
\end{equation} 
with $g^A_N$ the axial X17 coupling to the nucleon $N$. 
The photon-nucleon interaction is defined in terms of the Dirac and Pauli form factors,
\begin{align}
    &\Braket{N(p') | J^\mu_\gamma(0) | N(p)} \nonumber \\
    &\quad = -e \bar{u}(p') \left[ F_1(q^2) \gamma^\mu + \frac{i}{2m_N} F_2(q^2) \sigma^{\mu\nu} q_\nu \right] u(p),
\end{align}
where $q = p' - p$, $J^\mu_\gamma$ is the electromagnetic current operator, and $e > 0$ denotes the electric charge unit.
As the momentum transfer in the considered transition is very small, it is safe to approximate the form factors
with their value at $q^2 = 0$. That is, $F_1(0) = Q_N$, with $Q_p = 1$ ($Q_n = 0$) for proton (neutron) respectively.

The corresponding non-relativistic expansion of the matrix elements of $J_X^\mu$ and $J_\gamma^\mu$ in the long-wavelength limit gives rise to the dipole operators~\cite{Barducci:2022lqd}
\begin{equation}
    \hat{T}^\text{mag}_{1M} = \frac{i}{3 \sqrt{2}} g^A_N |\vb{k}_X| \hat{D}_{1M}, \quad \text{and} \quad
    \hat{T}^\text{el}_{1M} = \frac{\sqrt{2}}{3}e Q_N E_\gamma \hat{d}_{1M},
\end{equation}
respectively, with corresponding single-particle operators
\begin{align}
    D_{1M} &= \sqrt{\frac{3}{4\pi}} \left(\vb{r} \times \vb{\sigma}\right) \cdot \vh{e}_M = -i \sqrt{2} r \left[ \vb{Y}_1 \vb{\sigma} \right]_{1M}, \label{eq:mult1}\\
    d_{1M} &= \sqrt{\frac{3}{4\pi}} \left( \vb{r} \cdot \vh{e}_M \right) = r \left[ \vb{Y}_1 \vb{1} \right]_{1M} , \label{eq:mult2}
\end{align}
where $\hat{e}_M$, $M = \pm 1, 0$, are the spherical basis vectors, $Y_{LM}$ are the spherical harmonics and $\vb{r}$ is the position vector. Here we have introduced the tensor product of two spherical tensors via Clebsch-Gordan coefficients,
\begin{equation}
    U_{LM} = \left[ \vb{T}_{L_1} \vb{S}_{L_2} \right]_{LM} 
    = \sum_{M_1, M_2} \cg{L_1}{M_1}{L_2}{M_2}{L}{M} T_{L_1 M_1} S_{L_2 M_2},
\end{equation} 
and have used that for two rank-1 spherical tensors,
\begin{equation}
    \left( \vb{T} \times \vb{S} \right) \cdot \vh{e}_M = - i \sqrt{2} \left[ \vb{T}_1 \vb{S}_1 \right]_{1M}.
\end{equation}

It may be shown that for a spherical tensor in spin and isospin space,
\begin{align}
    &\redmel{\alpha_f; J_f T_f}{\vh{O}_L^T}{\alpha_i; J_i T_i} \nonumber \\
    &\quad = \hat{L}^{-1} \hat{T}^{-1}
    \sum_{a, b} \redmel{a}{\vb{O}^T_L}{b} \redmel{\alpha_f; J_f T_f}{\left[ c^\dagger_a \tilde{c}_b \right]_L^T}{\alpha_i; J_i T_i} \label{eq:redmelop},
\end{align}
where $c^\dagger_\alpha$ and $c_\beta$ are the single-nucleon creation and annihilation operators satisfying the conventional anti-commutation relations \cite{Suhonen:2007vjh}. We use the notation $\hat{j} = \sqrt{2j + 1}$. The corresponding hole operator is given by,
\begin{equation}\label{eq:holeop}
    \tilde{c}_\alpha = (-1)^{j_\alpha + 1/2 + m_{\alpha} + m_{t_\alpha}} c_{-\alpha}, \quad -\alpha = \left\{a, -m_\alpha, -m_{t_\alpha}\right\},
\end{equation}
with $a = n\ell sjt$, with $t = s = 1/2$, where  
we work in the coupled harmonic-oscillator basis $\Ket{n \ell s j tm m_t}$ with harmonic oscillator parameter \cite{Tsaran:2023qjx, DeVries:1987atn},
\begin{equation}
    a = \left( \frac{\hbar}{m\omega} \right)^{1/2} = 1.63 \text{ fm}.
\end{equation}
 
Under the assumption that the state $\Ket{{}^{12}\text{C}(17.23)}$ is predominantly a $2s_{1/2}1p^{-1}_{3/2}$ single particle-hole excitation we can write down,
\begin{align}
    \Ket{{}^{12}\text{C}(17.23)} &= \Ket{2s_{1/2}1p^{-1}_{3/2}; 1M 1 M_T} \nonumber \\
    &= \left[ c^\dagger_{2s_{1/2}} \tilde{c}_{1p_{3/2}} \right]^{1M_T}_{1M} \Ket{{}^{12}\text{C}(\text{g.s.})}.
\end{align}
A short calculation shows that the doubly-reduced one-body transition density for a 1p1h state decaying to the ground state is given by \cite{Suhonen:2007vjh},
\begin{equation}
    \redmel{0}{\left[ c^\dagger_a \tilde{c}_b \right]^T_L}{a_i b^{-1}_i; J_i T_i} = 
    \delta_{LJ_i} \delta_{TT_i} \delta_{ab_i} \delta_{ba_i} (-1)^{j_{\alpha_i} - j_{\beta_i} + J_i + T_i} \hat{J}_i \hat{T}_i,
\end{equation}
leading to
\begin{equation}
    \redmel{{}^{12}\text{C}(\text{g.s.})}{\vh{O}^T_L}{{}^{12}\text{C}(17.23)} 
    = - \delta_{L1} \delta_{T1} \redmel{1p_{3/2}}{\vb{O}^T_L}{2s_{1/2}}.
    \label{eq:Cmatrixele}
\end{equation}

Lastly, we compute the single-particle matrix elements of $\vb{D}_1$ and $\vb{d}_1$ on the {\it rhs} of Eq.~(\ref{eq:Cmatrixele}).
As the considered decay is an isovector transition, the couplings will enter as $g_p - g_n$ and $Q_p - Q_n$ respectively. We denote the radial matrix element by $\mathcal{R}^{(1)}_{1p,2s}$, where we have defined: 
%\begin{equation*}
%    \frac{(g_p - g_n)}{2} \mathcal{R}^{(1)}_{1p,2s} \sqrt{6}, \quad \text{and} \quad \frac{e (Q_p - Q_n)}{2} \mathcal{R}^{(1)}_{1p,2s} \sqrt{6}, 
%\end{equation*}  
\begin{equation}
    \mathcal{R}^{(\lambda)}_{n_a \ell_a, n_b \ell_b} = \int_0^\infty \mathrm{d}r \, r^{2} R^\ast_{n_a \ell_a}(r) r^\lambda R_{n_b \ell_b}(r), 
\end{equation}
with $R_{n\ell}(r)$ the normalized radial wave functions. For the harmonic oscillator basis wave functions one has:
\begin{equation}
    \mathcal{R}^{(1)}_{1p,2s} = -a. 
\end{equation}
Furthermore, let $\vb{T}_{L_1}$ and $\vb{S}_{L_2}$ be two commuting spherical tensor operators acting on the bases $\Ket{j_1 m_1}$ and $\Ket{j_2 m_2}$, respectively. Then, matrix elements in the coupled basis $\Ket{j_1 j_2 jm}$ can be expressed using the Wigner 9$j$ symbol as,
\begin{align}
    \redmel{j_1j_2j}{\left[ \vb{T}_{L_1} \vb{S}_{L_2} \right]_{L}}{j'_1 j'_2 j'}
    &= \hat{j} \hat{L} \hat{j}' \wnj{j_1}{j_2}{j}{j'_1}{j'_2}{j'}{L_1}{L_2}{L} \nonumber \\
    &\quad \times \redmel{j_1}{\vb{T}_{L_1}}{j'_1} \redmel{j_2}{\vb{S}_{L_2}}{j'_2}. \label{eq:split}
\end{align}
Using Eq.~\eqref{eq:split} and $\vb{r} = r \vh{r}$ one obtains~\cite{Suhonen:2007vjh}
\begin{equation}
    \redmel{p_{3/2}}{\left( \vh{r} \times \vb{\sigma} \right)}{s_{1/2}} = -i \frac{2}{\sqrt{3}}, \quad 
    \redmel{p_{3/2}}{ \vh{r} }{s_{1/2}} = \frac{2}{\sqrt{3}}.
\end{equation}
Therefore, together with $\redmel{\tfrac{1}{2}}{\vb{\tau}}{\tfrac{1}{2}} = \sqrt{6}$,
\begin{align}
    \redmel{1p_{3/2}}{\vb{D}_1 \vb{\tau}/2 }{2s_{1/2}} &= -i \sqrt{2} \sqrt{\frac{3}{4\pi}}  \mathcal{R}^{(1)}_{1p,2s}, \\
    \quad \redmel{1p_{3/2}}{\vb{d}_1 \vb{\tau}/2 }{2s_{1/2}} &= \sqrt{2} \sqrt{\frac{3}{4\pi}} \mathcal{R}^{(1)}_{1p,2s}.
\end{align}

Putting everything together yields the desired decay rates,
\begin{align}
    &\Gamma\left[ {}^{12}\text{C}(17.23) \to {}^{12}\text{C}(\text{g.s.}) + \text{X17}\right] = \frac{|\vb{k}_X|^3}{162\pi} (g^A_p - g^A_n)^2 \left|\mathcal{R}^{(1)}_{1p,2s}\right|^2, \label{eq:x17dec} \\
    &\Gamma\left[ {}^{12}\text{C}(17.23) \to {}^{12}\text{C}(\text{g.s.}) + \gamma \right] =  \frac{2e^2 E_\gamma^3}{81\pi}(Q_p - Q_n)^2 \left|\mathcal{R}^{(1)}_{1p,2s}\right|^2. \label{eq:emdec} 
\end{align}
Their ratio is independent of the radial wave function,
\begin{equation}\label{eq:ratio}
    \frac{\Gamma_X}{\Gamma_\gamma} = \frac{1}{4}\left[ 1 - \left( \frac{m_X}{\Delta E} \right)^2 \right]^{3/2} \frac{1}{e^2} \left( \frac{g^A_p - g^A_n}{Q_p - Q_n} \right)^2,
\end{equation}
where $\Delta E = 17.23$ MeV.

\section{Discussion}\label{sec:disc}

As mentioned in the introduction, the works of Refs.~\cite{VINHMAU196289, GOSWAMI1963294, GILLET1964321}, as well as the later works of  Refs.~\cite{PhysRevC.1.833, PhysRev.133.B849}, all consistently indicate that the $^{12}\text{C}(17.23)$ excited state is qualitatively well represented by a $2s_{1/2} 1p_{3/2}^{-1}$  particle-hole shell-model state, with only minor 2p2h contributions \cite{Grecksch1981}. One-particle-one-hole models typically offer a qualitative depiction of the spectrum and relative decay strengths, but tend to overestimate their absolute magnitudes. To estimate the theoretical uncertainty in our calculation, we first turn to the electromagnetic decay of ${}^{12}$C(17.23) to the ground state. Past calculations of inelastic electron scattering \cite{PhysRev.133.B849, PhysRev.139.B1217, Proca:1968dti, PhysRevC.1.833, Yamaguchi:1971ua} or semileptonic weak interactions in ${}^{12}$C \cite{PhysRevC.6.1911, PhysRevC.6.719} using the 1p1h framework require the reduction of the calculated cross sections by factors around two to five to match experimental results. Likewise, a calculation similar to ours, where the electromagnetic decay strength of ${}^{12}$C(16.11) to the ground state is calculated assuming the state is a pure particle-hole excitation, overestimates the result by a factor of four when compared to experiment \cite{Friebel:1978bug}. Consequently, we anticipate needing a similar reduction factor. Indeed, using Eq.~\eqref{eq:emdec} we find,
\begin{equation}
    \Gamma\left[ {}^{12}\text{C}(17.23) \to {}^{12}\text{C}(\text{g.s.}) + \gamma \right]
    \approx 251 \text{ eV}.
\end{equation}
This should be compared to the experimental value $\Gamma_\gamma^\text{exp}=44$~eV \cite{Kelley:2017qgh, Segel:1965zz}. Therefore, we require a reduction factor $\approx 5.4$. Such a factor is on the high end, but not entirely unreasonable given the simplifying nature of our approximations. Note that no uncertainty is given with the experimental values. 

In passing we should also mention that in the determination of the experimental decay width Segel \textit{et al.} found that a value $\Gamma_\gamma^\text{exp}=290$~eV may also describe the data \cite{Segel:1965zz}. However, based on comparison to other experiments, they give the smaller result as a preferred value. With the inclusion of a reduction factor our result also agrees with the smaller value. Nevertheless, it is worth mentioning that more recent fits to ${}^{12}$C occasionally still point to the larger solution \cite{PhysRevC.109.014619}. In this work we keep $\Gamma_\gamma = 44$ eV. However, as discussed below, replacing $\Gamma_\gamma$ with 290 eV does not alter our main conclusions. 

\begin{figure}[t!]
    \centering
    \includegraphics[width=8.8cm]{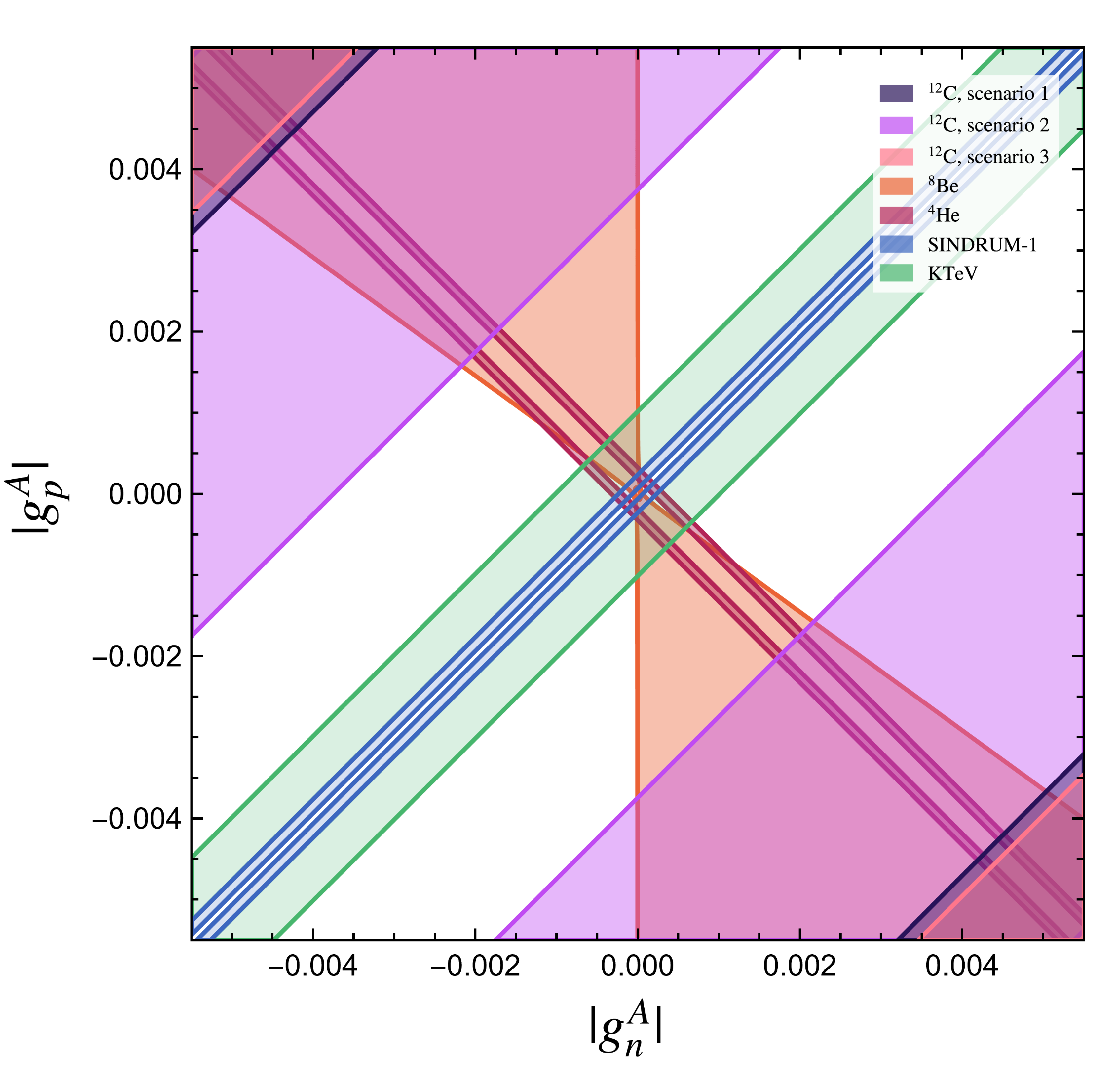}
    \caption{Compatibility regions for X17. Colored regions indicate combinations of the axial X17-nucleon couplings consistent at $1\sigma$ with the ATOMKI data $\Gamma_X/\Gamma_\gamma$ of ${}^8$Be \cite{Krasznahorkay:2015iga, Krasznahorkay:2015ijz} (orange), ${}^4$He \cite{Krasznahorkay:2021joi} (red) and ${}^{12}$C \cite{Krasznahorkay:2022pxs}, as well as preexisting constraints from $\pi^0 \to e^+ e^-$ of the KTeV collaboration (green) \cite{KTeV:2006pwx} and from $\pi^+ \to e^+ \nu_e e^+ e^-$ of SINDRUM-I (blue) \cite{SINDRUM:1989qan}. For the latter two the top bands correspond to the choice of a positive X17 axial electron coupling, while the bottom bands correspond to the negative choice. Further details for non-carbon decays and preexisting bounds can be found in Refs.~\cite{Barducci:2022lqd, Hostert:2023tkg, Mommers:2024eeq}. For carbon we consider three scenarios in which we 1) take the experimental value for $\Gamma_\gamma$ \cite{Segel:1965zz} and apply a reduction factor of 5.4 to the 1p1h shell-model estimate of $\Gamma_X$ (dark purple), 2) same as 1 but without the reduction factor (light purple), and 3) where the ratio  $\Gamma_\gamma / \Gamma_X$ is determined from the shell-model estimate (pink, nearly coinciding with scenario 1 bands). Results either point to that the axial-vector interpretation of the ATOMKI data is in tension with preexisting bounds, or to large corrections to the 1p1h shell-model estimate.}
    \label{fig:compat}
\end{figure}

Let us now apply our results to the ATOMKI ${}^{12}$C experiment, which measured the decay ratio as~\cite{Krasznahorkay:2022pxs}:
\begin{equation}
\label{eq:ratioatomki}
    \left( \frac{\Gamma_X}{\Gamma_\gamma} \right)_\text{ATOMKI} = 3.6(3) \times 10^{-6}.
\end{equation}
Analogously as was done in Ref.~\cite{Barducci:2022lqd} for the beryllium and helium decays, we use Eqs.~\eqref{eq:x17dec} and \eqref{eq:emdec} to scan the $g^A_p$-$g^A_n$ parameter space for the case of the $^{12}$C decay. We aim to find regions where the proton and neutron couplings are compatible with $(\Gamma_X / \Gamma_\gamma)_\text{ATOMKI}$ at the $1\sigma$ level. These $1\sigma$ compatibility regions are shown in Fig.~\ref{fig:compat}. The previously-derived compatibility regions from beryllium and helium decays are shown in orange and red, respectively and expressions thereof, which we also use here, may be found in Ref.~\cite{Barducci:2022lqd}. For carbon, we consider three scenarios: scenario 1) where we fix $\Gamma_\gamma = 44$ eV and use Eq.~\eqref{eq:x17dec} for $\Gamma_X$ with a reduction factor of 5.4, scenario 2) same as scenario 1) but without the reduction factor, and scenario 3) where we use the ratio of Eq.~\eqref{eq:ratio} fixed to the measured ATOMKI value of Eq.~(\ref{eq:ratioatomki}). In Fig.~\ref{fig:compat} scenarios 1), 2) and 3) are shown in dark purple, light purple, and pink, respectively. The different regions are also summarized in Table~\ref{tab:coupling}. The bands corresponding to scenario 3) are much closer to those of scenario 1) than 2), which is to be expected as in scenario 3) any reduction factors cancel. In Fig.~\ref{fig:compat} we also show existing constraints on the axial-vector nucleon coupling. Limits from the decay $\pi^0 \to e^+ e^-$ (KTeV anomaly) \cite{KTeV:2006pwx} are shown in green, and limits from $\pi^+ \to e^+ \nu_e e^+ e^-$ (SINDRUM-I) \cite{SINDRUM:1989qan} are shown in blue \cite{Hostert:2023tkg}. Both these external constraints depend on the sign of the axial coupling of X17 to the electron. The top bands correspond to a positive sign choice and the bottom bands correspond to a negative sign choice. More details, derivations and expressions of these constraints are given in Refs.~\cite{Kozaczuk:2016nma, Barducci:2022lqd, Hostert:2023tkg, Mommers:2024eeq}.

\begin{table}[b]
\centering
\caption{Constraints on the nucleon couplings for an axial-vector X17, derived from ATOMKI data on ${}^{12}$C(17.23) decay to the ground state, $\Gamma_X / \Gamma_\gamma$, using a $2s_{1/2} 1p_{3/2}^{-1}$ shell-model state to describe ${}^{12}$C(17.23). In scenarios 1) and 2) we use the experimental value for the electromagnetic decay rate $\Gamma_\gamma$, with and without an empirical reduction factor for the shell-model result of the X17-decay rate, $\Gamma_X$, respectively. In scenario 3) we use shell-model results for the  ratio $\Gamma_\gamma / \Gamma_X$.
}
\label{tab:coupling}
\begin{tabular}{rl}
\hline \hline
Scenario & $|g^A_p - g^A_n|$                \\ \hline
1        & $(0.87$-$2.5) \times 10^{-2}$ \\
2        & $(0.37$-$1.1) \times 10^{-2}$ \\
3        & $(0.90$-$2.6) \times 10^{-2}$ \\ \hline \hline
\end{tabular}
\end{table}

Regardless of the scenario it is evident that, like the vector interpretation of X17, interpreting  the anomalies in the ATOMKI, KTeV and SINDRUM-1 data sets as all arising from a single axial-vector X17 leads to tensions. The relatively large axial couplings following from the ATOMKI $^{12}$C result are mainly due to the fact that the corresponding decay proceeds through a relative $l=1$ wave, proportional to the third power of the small relative momentum of X17, as evident from Eq.~(\ref{eq:x17dec}). 
The application of the reduction factor to the 1p1h shell model result only intensifies the tension. And, even though the theoretical uncertainty of the 1p1h estimate is large, as seen by the difference between scenarios 1) and 2) of Fig.~\ref{fig:compat}, it is not enough to reconcile the derived ATOMKI limits and the preexisting constraints. For example, to yield consistent results one would need to increase the theoretical result and not decrease it via a reduction factor in the ratio $\Gamma_X / \Gamma_\gamma$, which is contrary to the correction one would expect based on electromagnetic decay. Note, however, that, although there is tension with the KTeV and SINDRUM-1 constraints, strictly speaking the three compatibility regions derived from the ${}^8$Be, ${}^4$He and ${}^{12}$C ATOMKI data are not inherently in conflict with each other, and at $\geq 2\sigma$ both theoretical and experimental uncertainty is sufficiently large that no clear conclusion can be drawn. Nevertheless, in view of the ${}^{12}$C shell-model result, consistency at $1\sigma$ between the different data sets would require at least one of the nucleon couplings to be $\mathcal{O}\left(10^{-2} \right)$, implying that the up and down quark couplings would have to be of a similar size as well. Contingent upon the electron and muon couplings, such large coupling values could bring additional tension with rare $\eta$ decays, $\eta \to \mu^+ \mu^-$ \cite{Kozaczuk:2016nma} or, if the lepton couplings are vectorial, with atomic parity violation experiments \cite{Barducci:2022lqd}. Exactly defining the regions in parameter space where there would be tension in these cases strongly depends on the underlying UV-complete model for X17 and the exact value (not just the difference) of its proton, neutron and lepton couplings, and falls outside the scope of this Letter.

It may very well be that the 1p1h approximation breaks down. After all, in assuming ${}^{12}$C(17.23) is exclusively the state $2s_{1/2}1p^{-1}_{3/2}$ one may overlook additional nuclear effects. On other hand, despite the simplicity of the approximation, it has yielded remarkably good qualitative results in predicting the low-energy spectrum and inelastic electron scattering of ${}^{12}$C \cite{Yamaguchi:1971ua, VINHMAU196289}. And, as mentioned previously, higher-order corrections that have been calculated were found to be small \cite{Grecksch1981}. Given its previous successes, it stands to reason the 1p1h approximation should work here---at least as a first approximation---as well. 

To go beyond the approximation used above and to assess the quality of the computation will require performing a full shell-model calculation, which is beyond the scope of this Letter. As stated in the introduction, multiple X17 verification experiments are presently underway. Data analyses for CCPAC \cite{Azuelos:2022nbu} and PADME \cite{Darme:2022zfw} are expected to conclude in the near future. If a signal is detected, then our 1p1h calculation indicates that either the interpretation of the ATOMKI data sets and corresponding constraints in terms of an axial-vector X17 needs to be reexamined, or that a more comprehensive shell-model calculation of the matrix elements of the spin-dipole operator $\vh{D}_1$ is needed.

\section{Summary and outlook}\label{sec:outlook}

The ATOMKI collaboration’s recent findings suggest the existence of a new 17-MeV boson, X17, which may either be a vector or an axial vector. Subsequent analysis \cite{Barducci:2022lqd, Hostert:2023tkg} studied the vector-like scenario and has found strong tensions with existing constraints. To analyze a possible axial-vector interpretation, we studied many-body nuclear matrix elements of the spin-dipole operator between ${}^{12}$C(17.23) and the ground state of carbon, under the assumption that the ${}^{12}$C(17.23) state is well-approximated by the $2s_{1/2}1p^{-1}_{3/2}$ particle-hole excitation~\cite{VINHMAU196289}. Despite large theoretical uncertainty, we find that our shell-model estimate also indicates tension in the axial-vector scenario when including all existing constraints. Even though previous successes of the 1p1h approximation in low-lying states of ${}^{12}$C give confidence in the obtained estimates, it is warranted to revisit this calculation of the spin-dipole operator with a more comprehensive shell-model calculation.

\section*{Acknowledgements}
The authors thank V. Tsaran for helpful communications.

This work was supported by the Deutsche Forschungsgemeinschaft (DFG, German Research Foundation), in part through the Research Unit [Photon-photon interactions in the Standard Model and beyond, Project number 458854507 - FOR 5327], and in part through the Cluster of Excellence [Precision Physics, Fundamental Interactions, and Structure of Matter] (PRISMA$^+$ EXC 2118/1) within the German Excellence Strategy (Project ID 39083149).

\bibliographystyle{elsarticle-num} 
\bibliography{bib}

\end{document}